\documentclass[manuscript]{aastex63}

\usepackage{color}

\shortauthors{Kong et al.}

\begin{document}

\title{A model of double coronal hard X-ray sources in solar flares}

\correspondingauthor{Xiangliang Kong}
\email{kongx@sdu.edu.cn}

\author[0000-0003-1034-5857]{Xiangliang Kong}
\affiliation{School of Space Science and Physics, Institute of Space Sciences, Institute of Frontier and Interdisciplinary Science, Shandong University, Shandong, People's Republic of China}

\author[0000-0002-5983-104X]{Jing Ye}
\affiliation{Yunnan Observatories, Chinese Academy of Sciences, P.O. Box 110, Kunming, Yunnan 650216, People's Republic of China}

\author[0000-0002-0660-3350]{Bin Chen}
\affiliation{Center for Solar-Terrestrial Research, New Jersey Institute of Technology,
323 Dr. Martin Luther King Blvd, Newark, NJ 07102, USA}

\author[0000-0003-4315-3755]{Fan Guo}
\affiliation{Los Alamos National Laboratory, Los Alamos, NM 87545, USA}
\affiliation{New Mexico Consortium, 4200 West Jemez Rd, Los Alamos, NM 87544, USA}

\author[0000-0002-9258-4490]{Chengcai Shen}
\affiliation{Center for Astrophysics $\mid$ Harvard $\&$ Smithsonian, 60 Garden St, Cambridge, MA 02138, USA}

\author[0000-0001-5278-8029]{Xiaocan Li}
\affiliation{Department of Physics and Astronomy, Dartmouth College, Hanover, New Hampshire 03755, USA}

\author[0000-0002-0660-3350]{Sijie Yu}
\affiliation{Center for Solar-Terrestrial Research, New Jersey Institute of Technology,
323 Dr. Martin Luther King Blvd, Newark, NJ 07102, USA}

\author[0000-0001-6449-8838]{Yao Chen}
\affiliation{School of Space Science and Physics, Institute of Space Sciences, Institute of Frontier and Interdisciplinary Science, Shandong University, Shandong, People's Republic of China}

\author[0000-0002-0850-4233]{Joe Giacalone}
\affiliation{Department of Planetary Sciences, University of Arizona, Tucson, AZ 85721, USA}

\begin{abstract}
A number of double coronal X-ray sources have been observed during solar flares by \textit{RHESSI}, where the two sources reside at different sides of the inferred reconnection site. 
However, where and how are these X-ray-emitting electrons accelerated remains unclear.
Here we present the first model of the double coronal hard X-ray (HXR) sources, where electrons are accelerated by a pair of termination shocks driven by bi-directional fast reconnection outflows.
We model the acceleration and transport of electrons in the flare region by numerically solving the Parker transport equation using velocity and magnetic fields from the macroscopic magnetohydrodynamic simulation of a flux rope eruption.
We show that electrons can be efficiently accelerated by the termination shocks and high-energy electrons mainly concentrate around the two shocks.
The synthetic HXR emission images display two distinct sources extending to $>$100 keV below and above the reconnection region, with the upper source much fainter than the lower one. The HXR energy spectra of the two coronal sources show similar spectral slopes, consistent with the observations.
Our simulation results suggest that the flare termination shock can be a promising particle acceleration mechanism in explaining the double-source nonthermal emissions in solar flares.

\end{abstract}



\section{Introduction} \label{sec:intro}
Solar flares are the most powerful energy release phenomena in the solar system. During solar flares, magnetic energy accumulated in the solar corona is explosively released via magnetic reconnection and converted in the form of particle acceleration, plasma heating, and bulk flows \citep{shibata11,benz17}. An enormous number of electrons can be accelerated to high energies (up to tens of MeV) and contain a significant fraction of the dissipated magnetic energy \citep{lin76,emslie12,aschwanden17,warmuth20}.
High-energy electrons can further produce hard X-ray (HXR) emission via the bremsstrahlung mechanism and microwave emission via the gyrosynchrotron mechanism.
Therefore, nonthermal HXR and microwave emissions provide the prime diagnostics of the acceleration and transport of energetic electrons in solar flares \citep{holman11,white11}. Understanding the particle acceleration and radiation processes is a central problem in solar flares.

A coronal HXR source located well above thermal soft X-ray flare loops, i.e., the looptop source, was first reported by \citet{masuda94}. 
Similar events with nonthermal looptop sources in HXR and microwave have been observed since then \citep[e.g.,][]{melnikov02,petrosian02,krucker10,krucker14,oka15,gary18,chen20,ning19,yu20}.
The looptop nonthermal sources suggest that energy release and particle acceleration occur above the soft X-ray flaring loop.
Although magnetic reconnection is widely believed to play a key role in releasing magnetic energy, it remains unclear which acceleration mechanism dominates and can explain various observational signatures of particle energization in solar flares \citep{miller97,zharkova11}.
A fast-mode shock can form when reconnection outflows impinge upon reconnected closed loops, usually referred to as the flare termination shock (TS).
The flare TS was often invoked in the standard flare model to explain the looptop HXR source \citep{masuda94,shibata95,magara96}, and both theoretical and numerical models showed that the TS can be a promising candidate of acceleration mechanism \citep{tsuneta98,mann09,warmuth09,guo12,kong13,li13}.
The flare TS has long been predicted in numerical magnetohydrodynamic (MHD) simulations \citep[e.g.,][]{forbes86,magara96,yokoyama98,takasao15,takasao16,shen18,cai19,ruan20,wang21}, and observational evidence for the existence of a TS in multiple wavelengths has also been reported \citep[e.g.,][]{masuda94,aurass02,aurass04,mann09,warmuth09,chen15,chen19,polito18,cai19,cai22,luo21}.

In some flares, a second coronal source can be simultaneously observed above the looptop sources discussed earlier. Together the two coronal sources are known as double coronal X-ray sources \citep[e.g.,][]{sui03,sui04,veronig06,liu08,liu13,chen12,glesener12,su13,ning16,chen17}. 
A current sheet where magnetic reconnection takes place is believed to lie between the two coronal sources. 
Note that in most events the two coronal sources are observed below $\sim$30 keV, which may be due to limited sensitivity and dynamic range of the instruments at high energies.
\citet{chen12} showed that the upper coronal source can be detected up to $\sim$80 keV.
The similarity in HXR spectra suggests that the same particle acceleration mechanism is responsible for the two sources \citep{liu08,chen12}.
Previous MHD simulations have shown that a pair of TSs can be driven by bi-directional reconnection outflows, one above the flare loop and one below the flux rope \citep{chen00,magara00,nishida09,takahashi17,ye19,ye21,zhao20}.
However, compared to the lower TS at the looptop, observational evidence for the upper TS remains very rare.
\citet{aurass04} reported on the signatures of the upper and lower TSs as type II radio bursts appearing simultaneously but not harmonically related. 
To explain the radio observations in \citet{reiner00}, \citet{magara00} presented a theoretical model which suggested that metric type II radio bursts could be produced by the fast-mode shock formed below the erupting flux rope (i.e., the upper TS in flare model). 
We note that, some metric type II radio bursts that are not well associated with CME-driven shocks and attributed to the flare blast wave, may also be an indication of the upper TS \citep[e.g.,][]{magdalenic10}.

Due to the dramatic difference between the kinetic scales ($\sim$1-10 cm for electrons and $\sim$1-10 m for protons) and the macroscopic flare scale ($\sim 10^8$ m), numerical modeling of particle acceleration in solar flares has been very challenging. 
Recently, we have developed a macroscopic kinetic model for energetic particles that combines a particle acceleration and transport model with the MHD simulation of solar flares  \citep{kong19,kong20}. \citet{kong19} showed that electrons can be efficiently accelerated by the TS and a magnetic trap plays a critical role in both confining and accelerating electrons in the looptop region. The model for the first time numerically reproduces the necessary electron acceleration and spatial distribution for nonthermal looptop sources.
This is consistent with recent microwave spectral imaging observations where the microwave-emitting electrons concentrate near a local minimum of magnetic field at the looptop, referred to as a magnetic bottle \citep{chen20}.
\citet{kong20} further showed that the acceleration of electrons can be dynamically modulated as the plasmoids collide with the TS at the looptop and suggested that it may explain the quasi-periodic pulsations of nonthermal flare emissions.

In this paper, we present the first model of double coronal HXR sources in solar flares by numerically modeling the acceleration of electrons by a pair of flare TSs from MHD simulations and generating the synthetic HXR images that can be compared with flare observations.

\section{Numerical Methods} \label{sec:model}
We model the acceleration and transport of electrons by coupling the Parker transport equation \citep{parker65} with a macroscopic MHD simulation of the solar flare. This general framework has been discussed previously \citep{kong19,kong20,li18} and here we only provide a salient description.

\begin{figure}
\centering
\includegraphics[width=0.95\linewidth]{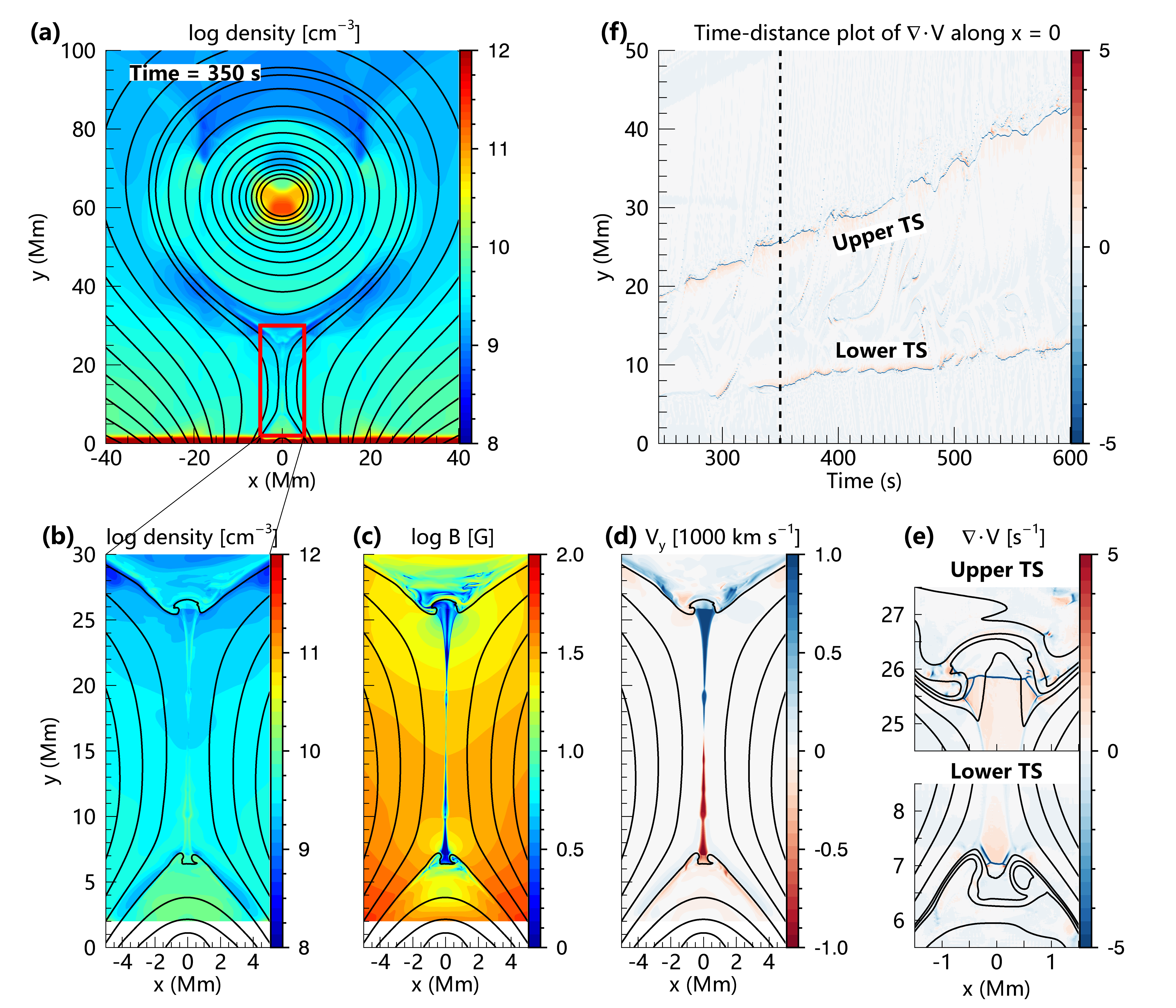}
\caption{
(a): Distribution of the plasma number density ($n_{the}$) and magnetic field configuration from MHD simulations at time 350 s. The black curves show the magnetic field lines.
(b)-(d): Distributions of the plasma number density ($n_{the}$), magnetic field strength ($B$), and plasma velocity ($V_y$) in the region taken for particle simulation (red box in panel (a)).
(e): Divergence of plasma flow velocity $\nabla \cdot \textbf{V}$ around the upper and lower TS regions.
(f): Time-distance plot of $\nabla \cdot \textbf{V}$ along $x$ = 0 in the period from 245 to 600 s. The vertical dashed line marks the time at 350 s.
}
\label{fig:mhd}
\end{figure}

For the flare MHD model, we use the 2.5D MHD simulation of a flux rope eruption with a Lundquist number of 1.18$\times 10^5$ in the vicinity of the current sheet as performed in \citet{ye19}. Interested readers are referred to \citet{ye19} for details of the MHD simulations.
A MHD snapshot at 350 s is shown in Figure \ref{fig:mhd}(a).
When the flux rope moves outward, a vertical current sheet forms below it and magnetic reconnection takes place.
Figures \ref{fig:mhd}(b)-(d) show the distributions of plasma number density ($n_{the}$), magnetic field strength ($B$), and plasma velocity ($V_y$) in the region taken for particle simulation, $x$ = [-5, 5] Mm and $y$ = [2, 30] Mm. 
The primary reconnection X-point is located at $y$ $\sim$16 Mm. Several plasmoids have been produced and move bidirectionally in the reconnection current sheet.
The reconnection outflows are super-Alfvenic in both directions and therefore can drive TSs \citep{ye19}.
Figure \ref{fig:mhd}(e) shows the divergence of plasma flow velocity $\nabla \cdot \textbf{V}$, in which the TSs are manifested by negative $\nabla \cdot \textbf{V}$ due to strong compression.
Figure \ref{fig:mhd}(f) shows the time-distance plot of $\nabla \cdot \textbf{V}$ along $x$ = 0 in the MHD time range of 245$-$600 s. During this period, the upper TS moves upward beneath the erupting flux rope with an average speed of $\sim$70 km s$^{-1}$, while the lower TS rises relatively slowly at $\sim$20 km s$^{-1}$.

Based on the velocity and magnetic fields from MHD simulations, we numerically model the acceleration and transport of electrons by solving the Parker transport equation \citep{parker65}.
It is achieved by integrating stochastic differential equations corresponding to the Fokker$-$Planck form of the transport equation using a large number of pseudo-particles \citep[e.g.,][]{zhang99,guo10,kong17,kong19,li18}.
With this approach we can examine how the electrons are accelerated and transport in the flare region.

The particle spatial diffusion coefficient describe particle transport in the magnetic field. The diffusion coefficient parallel to the mean magnetic field ($\kappa_{\parallel}$) can be evaluated according to the quasi-linear theory \citep{jokipii71,giacalone99}.
We assume that the magnetic turbulence is well developed and has a Kolmogorov power spectrum $P \propto k^{-5/3}$, then the resulting diffusion coefficient $\kappa_{\parallel} \propto p^{4/3}$ when the particle gyroradius is much smaller than the turbulence correlation length. 
Following our previous studies \citep{kong19,kong20}, we take $\kappa_{\parallel 0} = 2 \times 10^{15}$ cm$^2$ s$^{-1}$ for the electron injection energy $E_0$ = 5 keV.
We normalize the spatial length and plasma velocity by $L_0$ = 100 Mm and $V_0$ = 1000 km s$^{-1}$, then the normalization of diffusion coefficient $\kappa _0$ = $L_0 V_0$ = $10^{18}$ cm$^2$ s$^{-1}$, and the normalization of time is $t_0 = L_0/V_0$ = 100 s.
Therefore $\kappa_{\parallel 0}$ = 0.002 $\kappa_0$.
We also consider perpendicular diffusion and assume $\kappa_{\perp} / \kappa_\parallel$ = 0.05 as suggested in test-particle simulations \citep{giacalone99}.
In flare MHD simulations, we use adaptive mesh refinement in the region of interest and the grid size corresponds to a length of $\sim$7.8 km. The thickness of the TS is on the order of one grid cell, $\Delta_{sh} \sim$ 0.0001 $L_0$.
Therefore, the characteristic diffusion length at the shock is larger than the shock thickness, $L_{diff}$ = $\kappa_{nn} /V_{0} > \Delta_{sh}$, where $\kappa_{nn}$ is the diffusion coefficient in the shock normal direction. 
The time step is $\Delta t = 10^{-5}\ t_0$ at the injection energy, which ensures that pseudo-particles can see the shock transition.
Under these conditions, the particles are allowed to traverse the shock many times and be accelerated to high energies, which is the essence of diffusive shock acceleration.
We note that the particle mean free path in our simulation $\lambda_{\parallel} = 3 \kappa_{\parallel}/v \approx$ 20 km for 5 keV electrons, smaller than that deduced in some flares \citep[e.g.,][]{kontar14,musset18}. 
However, the turbulence properties and electron mean free path in the shock region remain unclear. We expect that magnetic turbulence around the shock region can reduce the electron mean free path. For simplicity, the diffusion coefficient is assumed to be a constant in different regions. 
To achieve efficient acceleration to hundreds of keV, our model requires the diffusion coefficient to be relatively small at the shock. Using a larger diffusion coefficient will reduce the particle acceleration rate.

In this study, we inject a total of 4.8$\times 10^6$ electrons with an initial energy of 5 keV uniformly in the simulation domain. 
In flare observations, both EUV spectroscopic and emission measure analysis have shown that the plasma in the current sheet and looptop region can be heated to $>$10 MK \citep{cheng18,polito18,warren18,chen21,cai22}. Therefore, heated plasma in the flaring region can provide seed population of electrons with a few keV. 
We have also tried different injection energies and a Maxwellian thermal distribution for 10 MK plasma and found that similar results can be obtained.
As a first step, we utilize a fixed MHD background as shown above and consider a characteristic acceleration time of 50 s. Although the height and shape of the TSs vary with time, here we only focus on the acceleration and transport of electrons in large scale.

The macroscopic particle model can provide spatially resolved distributions of high-energy electrons and can be used for further radiation modeling.
Based on the distributions of accelerated electrons and the plasma thermal density from the MHD simulation, we calculate HXR emission from energetic electrons.  
The grid numbers for calculating electron distributions from the particle simulation are $N_x = 100$ and $N_y = 300$, corresponding to a spatial resolution of $\sim$0.1 Mm.
To obtain the HXR emission maps at given energies, for each pixel, we calculate the thin-target bremsstrahlung X-ray spectrum by assuming the standard Bethe$-$Heitler cross-section using the Python package \texttt{sunxpsex}.
We use discrete electron energy spectra with the low-energy cutoff at 5 keV, the same as the lowest energy in the electron spectrum shown later.

Our current macroscopic acceleration model contains several limitations that are important to improve in the future. The Parker equation assumes a quasi-isotropic particle distribution which may not be valid for energetic electrons, especially when they leave the turbulent looptop region. In the next step, we plan to implement more generalized transport equations to consider more realistic electron distributions \citep[e.g.,][]{zank14,zhang17}. In addition, our current model does not include the feedback of energetic particles \citep{arnold21}, which may be important when reconnection and shocks accelerate a large fraction of particles into nonthermal distributions. Recent particle-in-cell kinetic simulations do suggest that these relevant acceleration processes are indeed quite efficient \citep{guo12,guo14,li19,zhang21}. Here, we mostly focus on explaining the morphology of the double coronal source features, and we will devote more effort to understanding detailed acceleration processes in the future.

\section{Simulation Results} \label{sec:result}

\begin{figure}
\centering
\includegraphics[width=0.9\linewidth]{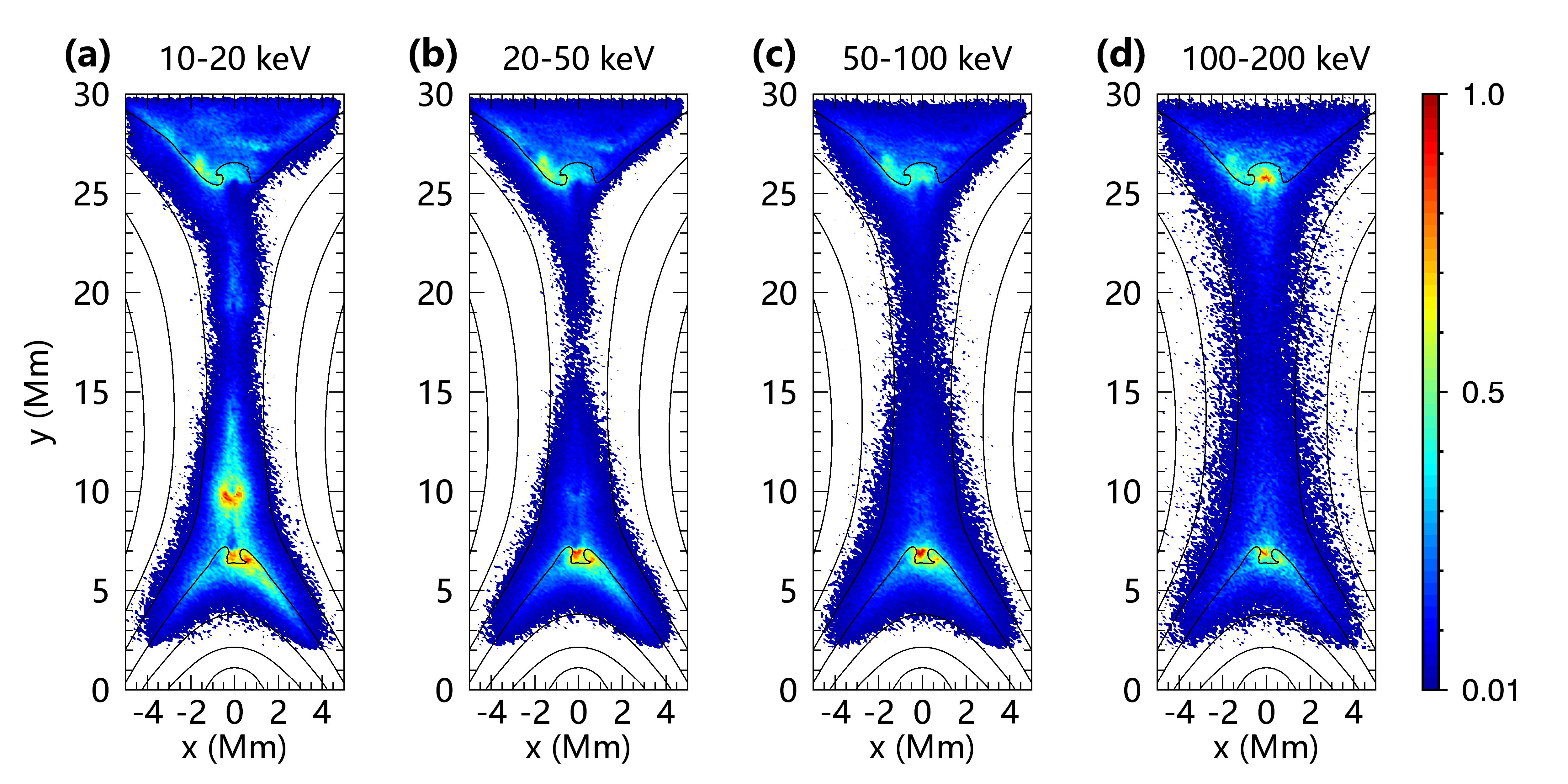}
\caption{
Spatial distributions of accelerated electrons at different energies. The values are normalized to the maximum of each individual image.
}
\label{fig:electron}
\end{figure}

Figure \ref{fig:electron} shows the spatial distributions of accelerated electrons at different energies, from $\sim$10 keV to $\sim$100 keV.
We observe two distinct sources in the simulation, one in the looptop region ($y \sim$7 Mm) and one below the flux rope ($y \sim$26 Mm). 
This indicates that both the upper and lower TSs can efficiently accelerate electrons to energies beyond 100 keV.
For the looptop source, the lower-energy electrons have a relatively broader distribution compared to that at higher energies. At 10$-$20 keV, an additional strong source appears above the looptop source at $y \sim$10 Mm, related to a plasmoid as shown in Figure 1. These low-energy electrons are probably accelerated by plasmoid-driven shock/compression and confined by closed field lines of the plasmoid \citep{li18}.
In contrast, electrons with higher energies concentrate close to the TS and in the concave-downward magnetic trap, agreeing with our previous studies \citep{kong19,kong20}.
For the upper source, at low energies (10$-$20 keV and 20$-$50 keV), most electrons cluster abnormally on the left side of the TS.
By examining the MHD data as shown in Figures 1(d) and 1(e), we suggest that low-energy electrons accelerated by the upper TS are advected to the ambient region on the left side with reconnecion outflows and confined by closed field lines.
Similar to the looptop source, the most energetic (100$-$200 keV) electrons in the upper coronal source are also mainly distributed around the TS in the magnetic trap.

\begin{figure}
\centering
\includegraphics[width=0.85\linewidth]{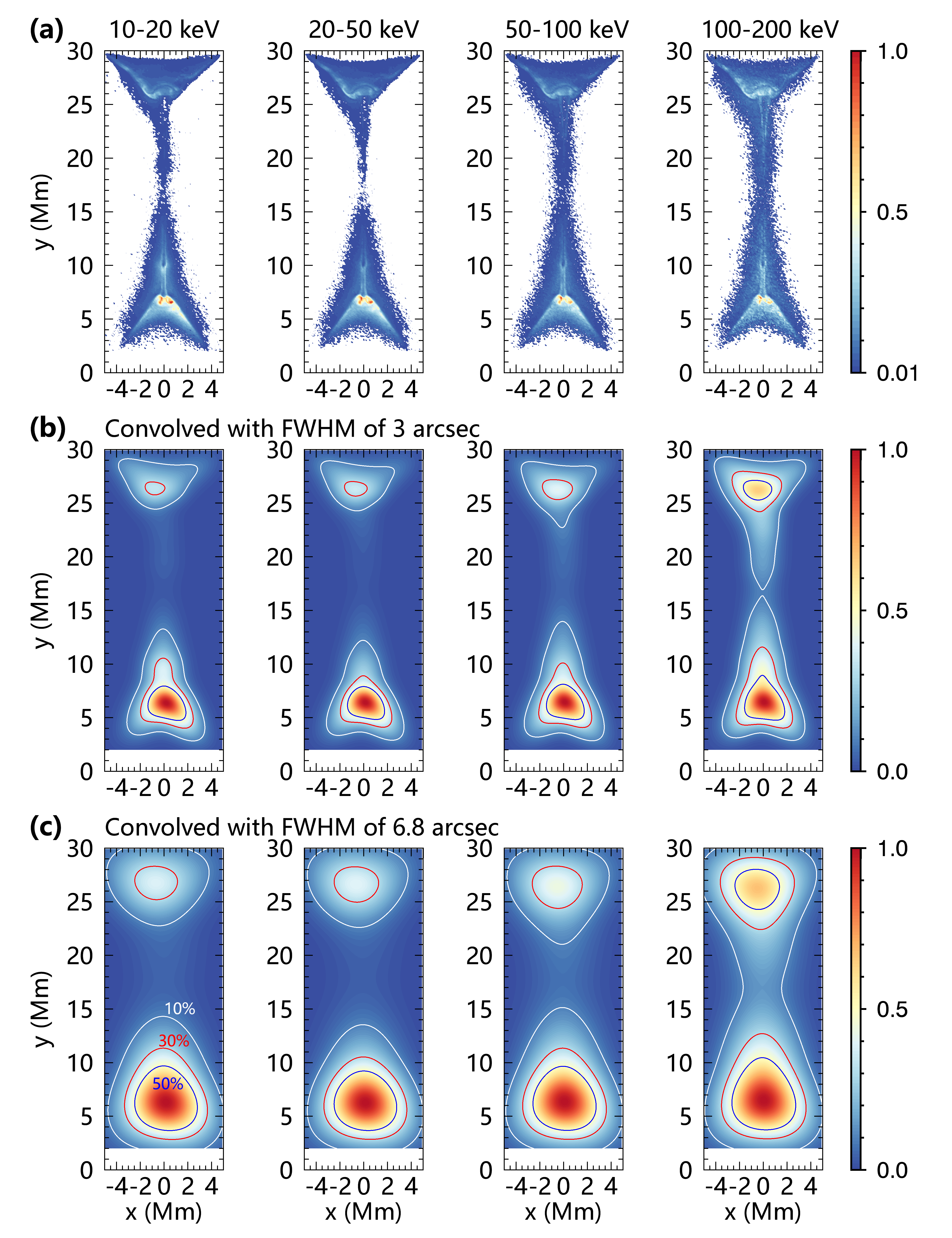}
\caption{
(a): HXR intensity maps for different photon energy ranges. (b)-(c): HXR images convolved with FWHM of 3 arcsec and 6.8 arcsec. The values are normalized to the maximum of each individual image and the contour levels in panels (b) and (c) are 10\% (white), 30\% (red), and 50\% (blue) of the maximum.
}
\label{fig:xray}
\end{figure}

By combining with the plasma thermal density from the MHD simulation, we calculate HXR emission from energetic electrons.  
Figure \ref{fig:xray}(a) shows the HXR intensity maps at different energy bands.
To compare our simulated results with flare observations, we convolve the HXR images with a Gaussian point-spread function with FWHM of 3$''$ and 6.8$''$, as shown in Figures 3(b) and 3(c), respectively.
A spatial resolution of 6.8$''$ is used to mimic the \textit{RHESSI} (detector 3) images, and a higher resolution of 3$''$ can provide predictions for future X-ray instruments.
The synthetic HXR images show two distinct HXR sources, above and below the reconnection region, with a distance of $\sim$20 Mm. 
However, the upper coronal source is much weaker than the lower source.
This is mainly because the plasma thermal density in the looptop is much higher,  $\sim$3 times that around the upper TS region (Figure 1(b)).
In addition, relatively more energetic electrons are distributed in the looptop.
In flare observations, due to limited dynamic range ($\sim$10:1 for \textit{RHESSI}), if the even brighter footpoint sources are present on the solar disk, the upper coronal source can hardly be detected.
This can explain why the upper HXR source is rarely observed compared to the looptop source, and if observed, the upper source was always weaker \citep[e.g.,][]{liu08,chen12}.
As shown in Figures 3(b), the high-resolution HXR images display a cusp-shaped looptop source, owing to faint emissions in the current sheet reconnection region ($\lesssim$10\% of the maximum intensity) and related to the plasmoid. It requires future instruments with high spatial resolution, high sensitivity and dynamic range to resolve these faint and fine HXR structures.

\begin{figure}
\centering
\includegraphics[width=0.9\linewidth]{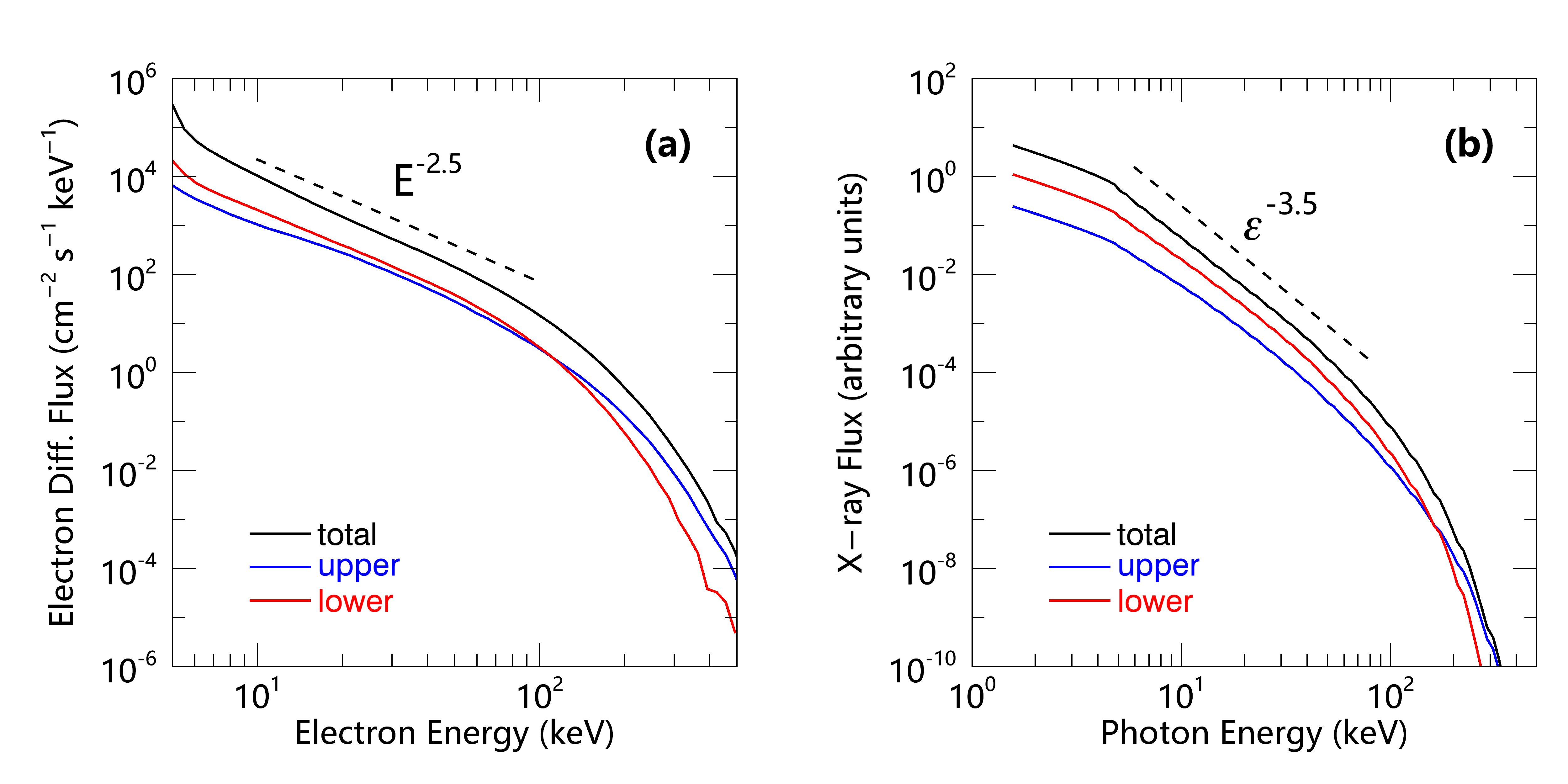}
\caption{
Energy spectra of accelerated electrons (a) and HXR emissions (b) integrated over the upper TS region (blue), the lower TS region (red), and the whole simulation domain (black). Note that the values are arbitrary and not normalized to the realistic values in flare observations.
}
\label{fig:spec}
\end{figure}

Figure \ref{fig:spec} shows the energy spectra of accelerated electrons and simulated HXR emissions. In addition to the spectra integrated over the whole simulation domain, we also compare the spectra of the upper and lower sources.
At energies below $\sim$100 keV, the three spectra show comparable spectral slopes.
The electron differential energy spectra are approximately a power-law distribution with a spectral index of $\delta$ = 2.5, consistent with our previous works \citep{kong19,kong20}.
We note that the electron spectrum of the upper source is slightly harder and has more electrons with energy $>$100 keV. Although the density compression ratio of the two TSs is similar, $\sim$2 on average, the length of the upper TS is longer.
In addition, the upper region on average is more strongly compressed. The mean of $\nabla \cdot \textbf{V}$ over the upper region (see Figure 1(e)) is twice as much as that of the lower region.
According to the Parker transport equation, the rate of energy gain is related to the magnitude of plasma compression, i.e., stronger compression will lead to more efficient acceleration \citep{li18,kong20}.
Therefore, in the upper region, the maximum energy that can be achieved for a given time is higher.

The X-ray spectra between 5 and 100 keV can be fitted with a power-law with a spectral index of $\gamma$ = 3.5. The relationship between the electron spectral index and the photon spectral index agrees with the thin-target X-ray emission model, $\gamma_{thin} = \delta + 1$ \citep{holman11}.
In accordance with the synthetic HXR images as discussed above, the integrated flux of the upper source is about 30\%-50\% of the looptop source at energies below $\sim$100 keV.
In the flare on 2003 Novermber 3, \citet{chen12} have shown that the spectral indices of the lower and upper coronal sources are both $\sim$3.8 for a similar energy range. \citet{liu08} fitted the X-ray spectra below $\sim$50 keV for a different flare event and also found that the nonthermal spectra of the two coronal sources have similar slopes (but with $\gamma$ ranging from 6 to 9) and the ratio between the upper and lower sources is $\sim$0.5.

\section{Conclusions and Discussion} \label{sec:conclusion}
\begin{figure}
\centering
\includegraphics[width=0.9\linewidth]{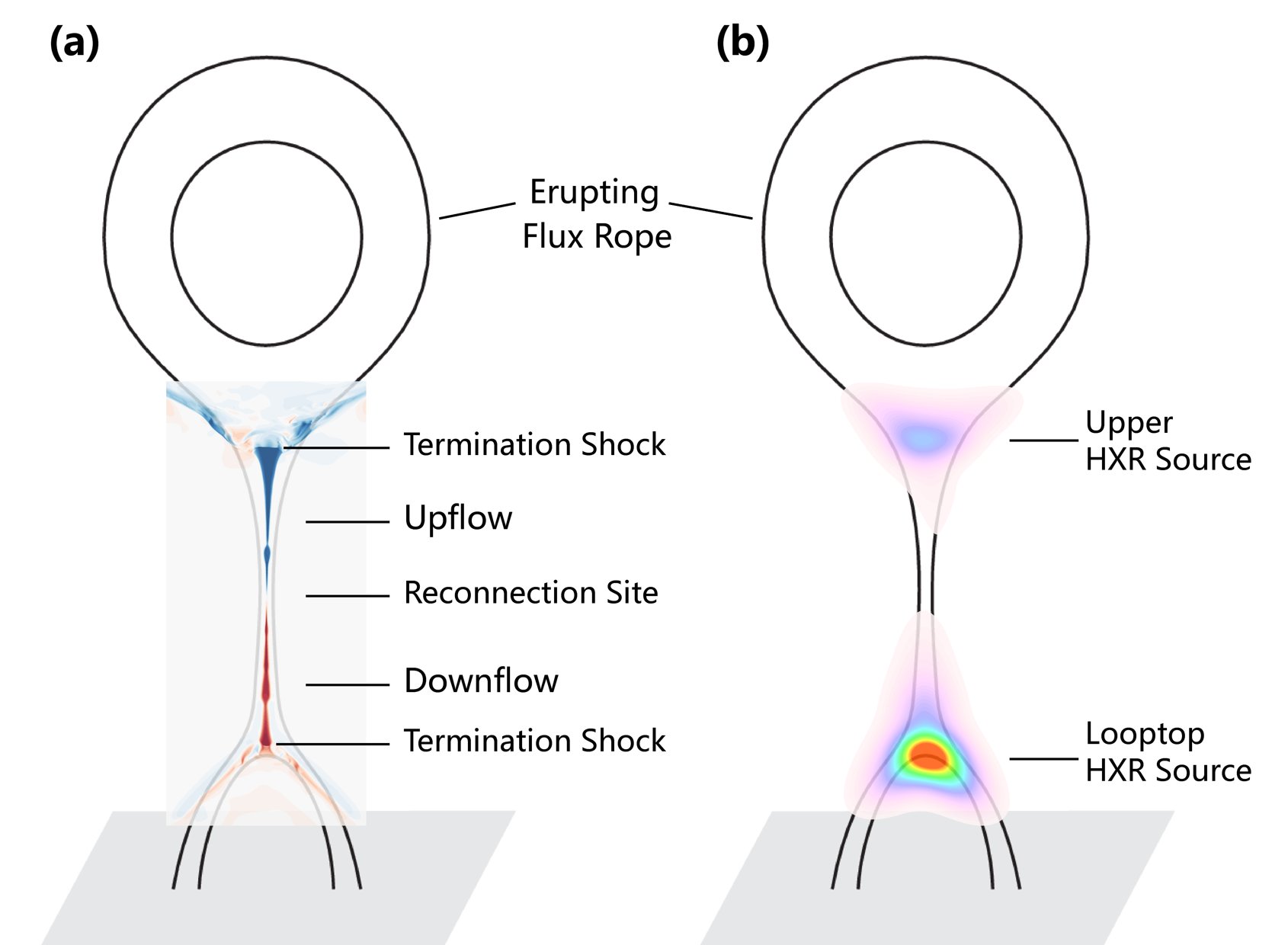}
\caption{
Schematics of the model of double coronal HXR sources. In an eruptive solar flare, a pair of TSs can form, one above the flare loop and one below the erupting flux rope. Electrons are accelerated by the TSs and concentrate around the two shocks, which can produce the double HXR sources.
}
\label{fig:cartoon}
\end{figure}

In this paper, we present the first model of double coronal HXR sources in an eruptive solar flare (see the schematics in Figure \ref{fig:cartoon}). Following our previous studies \citep{kong19,kong20}, we use a macroscopic kinetic model by numerically solving the Parker transport equation based on the MHD simulation of a flux rope eruption.
In the flare MHD simulation, fast bi-directional reconnection outflows can drive a pair of TSs \citep{ye19}.
Our particle simulations show that the two TSs can efficiently accelerate electrons to beyond 100 keV.
To compare with flare observations, we generate synthetic HXR emission images by calculating the thin-target bremsstrahlung X-ray emissions.
Two distinct HXR sources are produced, one at the looptop and the other below the flux rope, while the upper source is much fainter than the lower source.
At energies between 5-100 keV, the energy spectra of both sources show similar spectral slopes and can be fitted with a power-law function.
These simulation results are consistent with the nonthermal HXR emissions of double coronal sources observed in solar flares \citep{chen12}.

In flare observations, it has been shown that coronal X-ray sources at higher energies are located closer to the reconnection site \citep[e.g.,][]{sui03,liu08}. However, we note that this energy-dependence trend in locations is mostly found in thermal emissions below 20 keV, and it remains uncertain for the nonthermal emissions at higher energies \citep{liu08,chen12}. In our simulation, such an energy dependence is not obvious. In our previous work \citep{kong19}, the electron distribution shows that the higher the electron energy, the source size is smaller and the source is located closer the termination shock (in higher location). In our future work, we will further examine this issue by using time-dependent MHD data and consider other mechanisms that can affect the acceleration and transport of energetic electrons, e.g., magnetic mirroring and Coulomb collision.

The high-resolution synthetic HXR images suggest that if the future X-ray instruments can achieve both high dynamic range ($>$50:1) and spatial resolution ($<$2$''$), some faint and fine structures, e.g., a cusp-shaped structure as observed in soft X-ray and EUV wavelengths, and separated sources related to the plasmoids, may be revealed. 
For example, FOXSI-4 \citep{buitrago21} and \citep{narukage19} with new focusing X-ray optics may be able to unveil faint coronal sources.
Hard X-ray Imager (HXI) onboard the ASO-S mission, scheduled to be launched in 2022, can achieve a spatial resolution of $\sim$3$''$ in the energy range of 30-200 keV, but the dynamic range is similar to \textit{RHESSI} \citep{gan19}.

Where and how are electrons energized has been a long-standing unsolved fundamental problem in solar flare studies. As discussed in previous studies \citep{liu08,chen12}, turbulence can be generated in the reconnection outflows and therefore stochastic acceleration by turbulence or plasma waves is a possible mechanism for explaining the coronal X-ray sources.
Our simulations suggest that the flare TS can be a promising candidate for acceleration mechanism, particularly in explaining the double coronal HXR sources. 
However, the observational evidence for the presence of the upper TS remains rare. To support the double-termination-shock model, in the future it is necessary to find more evidence in flare observations.
Since our macroscopic particle model enables detailed comparison with flare observations, it may have strong implications in interpreting nonthermal emissions in solar flares.


\acknowledgments
This work was supported by the National Natural Science Foundation of China under grants 11873036, 42074203 and 11790303 (11790300), the Young Elite Scientists Sponsorship Program by China Association for Science and Technology, and the Young Scholars Program of Shandong University, Weihai.
J.Y. is supported by NNSFC grant 12073073 and the grant associated with the Applied Basic Research of Yunnan Province 202101AT070018.
B.C. acknowledges support by NSF grant AST-1735405 to NJIT.
F.G. is supported by NSF grant AST-2109154 and DOE grant DE-SC0018240.
X.L. is supported by NSF grant AST-2107745.
The work was carried out at National Supercomputer Center in Guangzhou (TianHe-2).







\end{document}